# Beyond Anthropomorphism: Exploring the Roles of Perceived Non-humanity and Structural Similarity in Deep Self-Disclosure Toward Generative AI

Satoru Shibuya, Ph.D.
Graduate School of Business and Finance, Waseda University

Email: sat.shibuya@waseda.jp

Abstract

This study investigates deep self-disclosure toward generative AI by examining perceived non-humanity and structural similarity as psychological factors beyond anthropomorphism. Perceived non-humanity may reduce evaluation apprehension, whereas structural similarity refers to the perceived logical alignment between a user's thinking and AI responses. Using cross-sectional survey data from 2,400 participants collected in 2025, this study analyzed associations with both the occurrence and depth of self-disclosure. Logistic regression indicated that the group high in both perceptions (Segment D) showed a significantly higher likelihood of disclosure than the baseline group (Segment A; OR = 11.35). ANOVA further showed significant between-group differences in disclosure depth. The findings suggest that trust-related behavior in deep self-disclosure may involve factors other than anthropomorphic perception. Because the study is exploratory and based on self-reported survey data, the results should be interpreted as associative rather than causal, and future longitudinal or experimental research is needed.

## 1. Introduction

Since 2025, the rapid proliferation of generative artificial intelligence (AI), exemplified by models such as ChatGPT and Gemini, has brought about a qualitative transformation in human-computer interaction (HCI). A particularly noteworthy phenomenon is users engaging in "deep self-disclosure" toward AI—the sharing of personally or socially sensitive information that individuals often hesitate to disclose in face-to-face interpersonal interactions.

The dominant paradigm in traditional HCI research has long posited that "anthropomorphism"—making AI more human-like—is the key to building trust (Nass et al., 1994). However, this anthropomorphic model may not fully explain the paradox of why humans feel greater psychological safety with AI, which is inherently non-human, than they do in interpersonal relationships. On the contrary, research suggests that as an agent becomes more human-like, users may become more conscious of "social evaluation," potentially suppressing self-disclosure due to shame or fear of social sanctions.

To address this issue, a new framework is proposed for explaining self-disclosure toward generative AI, drawing on the classical "structure-mapping theory" of similarity perception. The central premise is that, while the perceived non-humanity of AI may reduce social evaluation apprehension, the perception of logical consistency, or "structural similarity," between the user's thought process and the AI's responses may foster cognitive trust. Taken together, these two factors are expected to be associated with deep self-disclosure. Within this framework, the former is positioned as a "buffer against evaluation apprehension," while the latter functions as a "mechanism for building cognitive trust."

The primary objective of this study is to clarify how the perception of AI's non-humanity and structural similarity relates to the occurrence and depth of deep self-disclosure. A secondary objective is to examine how the combination of these two factors varies across different age groups and genders.

## 2. Theoretical Framework and Hypotheses

### 2.1. Avoidance of Social Evaluation Apprehension and Benefits of Non-anthropomorphism

In traditional HCI research, the CASA (Computers Are Social Actors) paradigm proposed by Nass et al. (1994) demonstrated that humans unconsciously apply social rules to computers. Based on this insight, anthropomorphism—granting AI human-like appearances, voices, or empathetic responses—has long been considered the key to fostering user trust and self-disclosure.

However, in the context of deep self-disclosure, this "humanness" can paradoxically become a psychological barrier. In interpersonal communication, self-disclosure always carries the risk of rejection or negative evaluation. The more a user perceives an AI as a human-like entity, the more they may unconsciously become conscious of social evaluation, potentially leading to self-inhibition due to shame or fear of social sanctions. This study identifies a challenge in the increase of social evaluation apprehension associated with anthropomorphism.

In contrast, the non-human attributes of AI—as "emotionless computing machines"—can serve to release users from this evaluation apprehension. For contemporary individuals bound by social roles and norms, a non-human AI may reduce concerns about social evaluation and form a disclosure environment with a relatively lower psychological burden. This perspective complements traditional discussions of anthropomorphism regarding trust formation in AI.

### 2.2. Structure-Mapping Theory (SMT) and Building Cognitive Trust

Mere "non-humanity" serving as a "safe wall" is insufficient for building deep trust through dialogue. In this regard, Gentner's (1983) "Structure-mapping Theory" (SMT), a classic theory of similarity perception, is highly relevant. According to SMT, when humans judge the similarity between two entities, they place greater emphasis on "relational mappings" (structural consistency) between elements than on "attributes" (surface-level consistency) such as color or shape, thereby engaging in higher-order analogy.

Shibuya (2012) applied this theory to the context of consumer behavior, suggesting that in interactions between the sender and receiver of word-of-mouth information, structural consistency in thought processes—rather than similarity in attributes—is associated with the credibility and confirmation of information. Extending this insight to interactions with generative AI, trust may depend not on the AI being similar to the user in terms of attributes, but on the extent to which the logical structure of the user's concerns and thoughts is accurately traced and reconstructed during dialogue with the AI. When users perceive that the AI is tracing their thought processes or logical structures to a certain degree, they are more likely to experience a sense of "being understood." Here, this perception is conceptualized as "perceived structural similarity." In other words, structural similarity refers not to appearance-based or attribute-based similarity, but to similarity in logical relations grounded in Gentner's (1983) Structure-mapping Theory.

The trust-building role of perceived structural similarity can also be reinforced theoretically from the perspective of Norman's (1988) concept of mental models, a core idea in Human-Computer Interaction (HCI) research.

According to Norman (1988), users construct internal models to predict and explain a system's behavior through interaction with it. In this context, structural similarity can be interpreted as a state in which the user's mental model of their own thought processes is closely aligned with the logical structure presented by the AI. When a structure mapping based on Gentner's (1983) Structure-mapping Theory is established, the user perceives that the AI is accurately tracing their line of thought, thereby generating predictability and cognitive trust toward the system. This form of cognitive trust, independent of emotional pathways such as anthropomorphism, may function as a powerful cognitive incentive that helps users overcome psychological barriers to self-disclosure.

While previous studies have indicated that the humanness of AI can enhance rapport and trust, they have not sufficiently examined negative aspects such as the arousal of evaluation concerns and shame. From this perspective, the present analysis reconsiders the conditions for self-disclosure by focusing on the combination of non-humanity and structural similarity.

It should be noted, however, that Structure-mapping Theory was originally developed to explain analogy and the mapping of relational structures between concepts. Accordingly, the current study does not assume that the AI's response itself possesses inherent structural similarity. Rather, it assumes a cognitive process in which the user perceives the AI's response by mapping it onto their own thought structure. In this sense, the SMT framework is employed to operationally and exploratorily examine the user's subjective perception of structural similarity in relation to AI responses.

### 2.3. Structural Non-humanity Model

The "Structural Non-humanity Model" proposed in this study integrates two independent mechanisms:

1. The Buffer Factor (Non-humanity): Ensures psychological safety by reducing social evaluation apprehension.

2. The Cognitive Trust Factor (Structural Similarity): Fosters cognitive trust through the alignment of logical structures.

When these two elements coexist, the AI is perceived as a non-judgmental yet logically consistent entity, which may be associated with deeper self-disclosure.

### 2.4. Hypotheses

Based on this framework, the following hypotheses were formulated:

H1: Perceived non-humanity of generative AI is positively associated with the depth of self-disclosure.

H2: Perceived structural similarity between a user’s thought process and AI responses is positively associated with the occurrence of self-disclosure.

## 3. Methods

### 3.1. Participants and Sampling

The present study is based on a secondary analysis of data from 2,400 respondents who completed a survey on generative AI use, as part of a larger research project involving 10,000 individuals. The dataset was provided by Hakuhodo DY Holdings in a fully anonymized format, ensuring that no personally identifiable information was accessible to the authors. In accordance with the guidelines of the authors' affiliated institution, formal ethics committee approval was not required for this secondary analysis of anonymized corporate data. Furthermore, informed consent—including permission for the use of data for research purposes—was obtained from all participants by the original data provider at the time of the survey. The original study was conducted under the provider's internal research governance and ethical review procedures.

The survey was conducted online in 2025 among men and women aged 18 to 69 across Japan. The sample was designed using quota sampling based on national population estimates from the Ministry of Internal Affairs and Communications, with quotas aligned by gender and 10-year age groups. Because the analysis relies on existing survey data, each construct was measured using pre-existing questionnaire items. Accordingly, certain limitations regarding construct validity should be acknowledged, and this analysis should be regarded as an initial and exploratory examination of the proposed framework.

**3.2. Operational Definitions of Variables**

To examine the "Structural Non-humanity Model" of AI, the primary variables were operationalized as follows:

1. Perceived Non-humanity (Independent Variable 1): To capture the degree to which users recognize the "non-human attributes" of AI as a resource for psychological safety, item Q3_17 ("Because the partner is not human, I don't have to worry about what they think") was used as a proxy indicator. Respondents who selected this item were operationally defined as the "group with usage motivations accompanied by the perception of non-humanity." This is a reconstructed measure that reinterprets an existing question from the perspective of usage motivation to approximately capture the exclusion of evaluation risks in interpersonal communication.

2. Perceived Structural Similarity (Independent Variable 2): To measure the degree to which the AI maps the user's own logical structure, items concerning relationship perception toward AI (Q25S1_16: "The logical structure of the talk is similar to mine," and Q25S1_10: "The thought process is similar to mine") were used. Respondents who answered "Very much applies" or "Somewhat applies" were classified as the "High Perceived Structural Similarity group." This is a composite measure reflecting a state in which the AI's output is structurally synchronized (aligned) with the user's own thought processes, moving beyond surface-level similarity of attributes.

3. Occurrence of Self-disclosure (Dependent Variable 1): To measure self-disclosure behavior, a question regarding specific consultation content (Q23_25: "I consulted about matters that I cannot tell others") was used. The presence or absence of this behavioral experience was defined as a binary variable (1 = experienced, 0 = not experienced).

4. Depth of Self-disclosure (Dependent Variable 2): In this study, to perform an exploratory verification not only of the "presence" but also of the "depth" of disclosure using secondary data, the depth of self-disclosure was defined as an operational index consisting of four items. Specifically, four items related to emotional venting, sense of trust, sense of security, and proximity to sensitive information were extracted from the 35 items in Q23, and a "Self-disclosure Depth Score" ranging from 0 to 4 was calculated by simply summing the number of selected items.

   In constructing the "Self-disclosure Depth Score," based on Social Penetration Theory (Altman & Taylor, 1973), four items considered to reflect the depth of self-disclosure were extracted from the 35 items in Q23 (Q23_5: Emotional venting; Q23_13: Sense of trust; Q23_15: Sense of security; Q23_25: Consultation regarding sensitive information). These items measure the disclosure of emotional and sensitive information beyond the mere transmission of facts, as well as the accompanying psychological commitment.

   Upon confirming the internal consistency of these four items, the Cronbach’s α coefficient was 0.662, which falls below the general threshold (0.70). For this reason, this index must be interpreted as an exploratory indicator that approximately captures the multifaceted aspects of self-disclosure. It should be noted that the item-total correlations all showed a certain level, suggesting that all four items provide relevant information for the construction of the index.

This index is a composite measure created by extracting items from an existing set of questions and is not a standardized existing scale itself. Therefore, this index should be interpreted as an operational indicator for approximately grasping the tendency toward deep self-disclosure, rather than as a psychological scale in the strict sense.

Future research should develop and validate a more robust multi-item scale for disclosure depth to improve measurement reliability.

### 3.3. Segment Classification and Analytical Approach

Participants were classified into four segments based on the combination of the two independent variables:

Segment A (Baseline): Perceives neither non-humanity nor structural similarity.

Segment B (Non-humanity Only): Perceives only non-humanity.

Segment C (Structural Similarity Only): Perceives only structural similarity.

Segment D (Structural Non-humanity): Perceives both non-humanity and structural similarity.

For the statistical analysis, a chi-square test was used to examine differences in self-disclosure rates (Dependent Variable 1) across the segments. In addition, logistic regression analysis was conducted as a supplementary analysis to control for covariates such as demographic attributes. Furthermore, a one-way analysis of variance (ANOVA) was performed to test differences in self-disclosure depth (Dependent Variable 2) among the four segments. In interpreting the results, attention was paid to the imbalance in sample sizes across groups, and both effect sizes and post-hoc comparisons were considered.

## 4. Results

### 4.1. Comparison of Self-disclosure Rates and Effect Sizes

To validate the effectiveness of the proposed model, the total sample ($N = 2{,}400$) was classified into four segments based on the presence or absence of "perceived non-humanity" and "perceived structural similarity." A significant association was observed between the segments and the occurrence of deep self-disclosure ($\chi^2(3) = 281.3$, $p < .001$).

Specifically, when calculating the odds ratios (ORs) for each segment relative to the Baseline (Segment A, 8.9%), the Structural Non-humanity (Segment D) segment exhibited the highest likelihood of disclosure (OR = 11.35, 95% CI [6.71, 19.20], $p < .001$). Even the Non-humanity

Only (Segment B) group showed a strong association (OR = 7.05, 95% CI [4.51, 11.01], $p < .001$), while the Structural Similarity Only (Segment C) group also remained significant (OR = 2.08, 95% CI [1.45, 2.97], $p < .001$). These findings suggest that the coexistence of both factors is associated with a markedly higher likelihood of self-disclosure occurrence.

Table 1 presents the percentages and odds ratios for deep self-disclosure across the four segments.

**Table 1** *Percentages and Odds Ratios for Deep Self-Disclosure across the Four Segments*

| Segments | *n* | Deep Self-disclosure | Deep S-D Rate (%) | OR | 95% CI | p-value |
|---|---|---|---|---|---|---|
| Segment A (Baseline) | 1,986 | 176 | 8.9 | 1 | (Reference) | - |
| Segment B (Non-humanity Only) | 91 | 37 | 40.7 | 7.05 | [4.51, 11.01] | <.001 |
| Segment C (Structural Similarity Only) | 262 | 44 | 16.8 | 2.08 | [1.45, 2.97] | <.001 |
| Segment D (Structural Non-humanity) | 61 | 32 | 52.5 | 11.35 | [6.71, 19.20] | <.001 |

Note. Segment A = low perceived non-humanity / low perceived structural similarity; Segment B = high perceived non-humanity only; Segment C = high perceived structural similarity only; Segment D = high perceived non-humanity / high perceived structural similarity. OR = odds ratio; CI = confidence interval. Segment A serves as the reference group.

**Figure 1** *Deep Self-Disclosure Rates across the Four Segments*

Percentage of Respondents (%)

|  | Rate |
|---|---|
| Segment A (Baseline) | 8.9% |
| Segment B (Non-humanity Only) | 40.7% |
| Segment C (Structural Similarity Only) | 16.8% |
| Segment D (Structural Non-humanity) | 52.5% |

Note. Segment A = low perceived non-humanity / low perceived structural similarity; Segment B = high perceived non-humanity only; Segment C = high perceived structural similarity only; Segment D = high perceived non-humanity / high perceived structural similarity. Values indicate deep self-disclosure rates (%).

Figure 1 shows the deep self-disclosure rates across the four segments. Regarding the odds ratios (ORs) for each segment relative to Segment A (Baseline: 8.9%), Segment B (Non-humanity Only) showed an OR of 7.05 (95% CI [4.51, 11.01]), while Segment C (Structural Similarity Only) showed an OR of 2.08 (95% CI [1.45, 2.97]). In Segment D (Structural Non-humanity), where both factors are integrated, the self-disclosure rate reached 52.5%, with an OR of 11.35 (95% CI [6.71, 19.20]).

These findings demonstrate a robust association between deep self-disclosure and the group characterized by high levels of both perceived non-humanity and perceived structural similarity compared to the baseline. It should be noted, however, that while these results indicate a strong link between specific perceptual patterns and self-disclosure behavior, they should be interpreted as statistical associations rather than direct causal relationships, as the magnitude of odds ratios can be sensitive to group definitions and reference standards. Figure 2 presents the odds ratios for deep self-disclosure with Segment A as the baseline.

**Figure 2** *Odds Ratios for Deep Self-Disclosure by Segment (Segment A as Reference Group)*

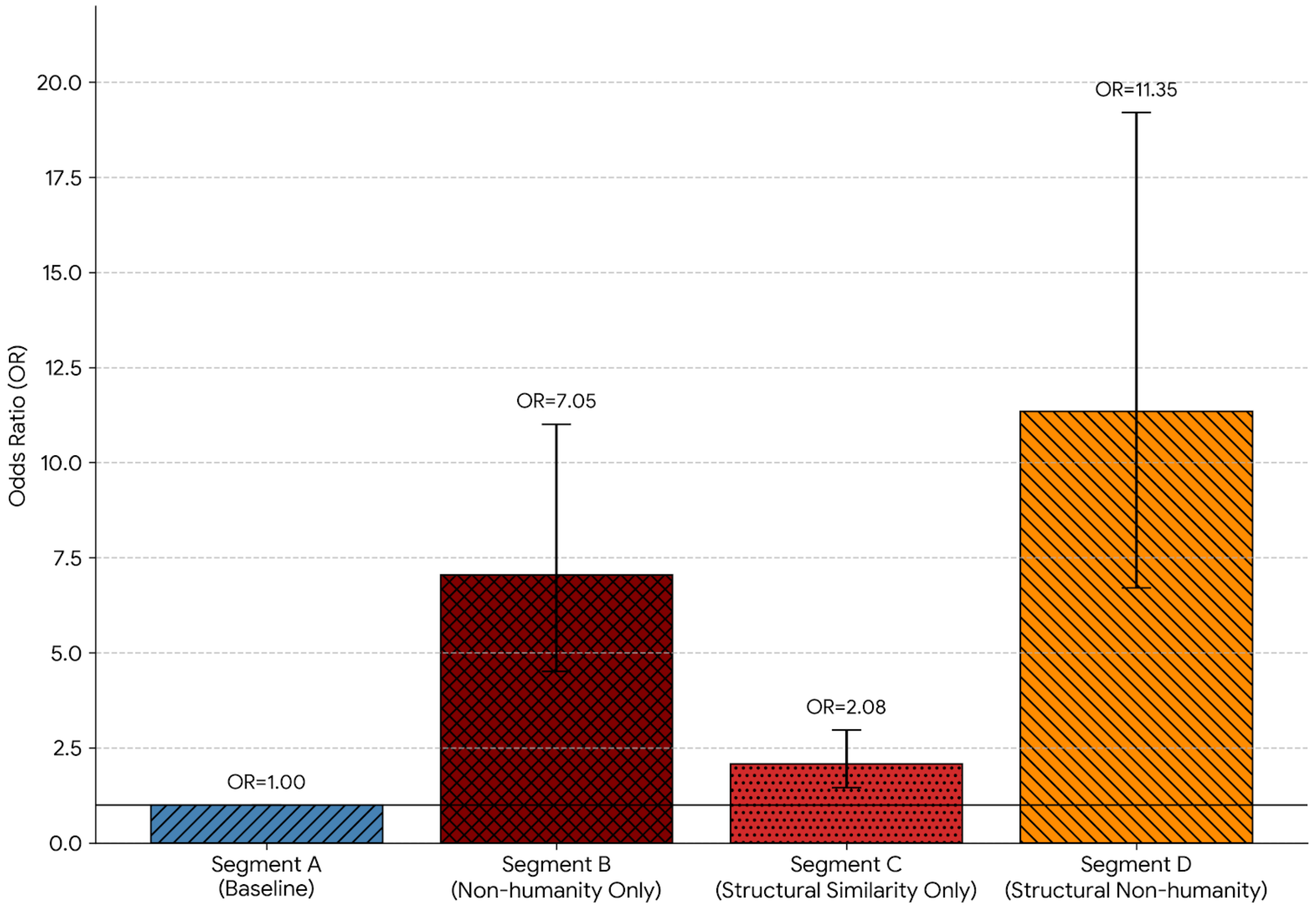

Note. Segment A = low perceived non-humanity / low perceived structural similarity; Segment B = high perceived non-humanity only; Segment C = high perceived structural similarity only; Segment D = high perceived non-humanity / high perceived structural similarity. OR = odds ratio; CI = confidence interval. Segment A serves as the reference group.

## 4.2. Comparison of Self-disclosure Depth and Effect Sizes

A one-way analysis of variance (ANOVA) revealed highly significant differences in self-disclosure depth across the four segments ($F(3, 2396) = 139.60$, $p < .001$). The effect size ($\eta^2 = .149$) exceeded the threshold for a "large" effect ($\eta^2 \geq .14$) in social sciences.

Table 2 presents the mean self-disclosure depth scores across the four segments and the results of the one-way ANOVA.

**Table 2** *Comparison of Self-Disclosure Depth Scores across the Four Segments*

| Segments | *n* | Mean | SD | *F* | *p* | $\eta^2$ |
|---|---|---|---|---|---|---|
| Segment A (Baseline) | 1,986 | 0.30a | 0.69 | 139.6 | <.001 | 0.149 |
| Segment B (Non-humanity Only) | 91 | 1.34b | 1.36 | | | |
| Segment C (Structural Similarity Only) | 262 | 0.69c | 1.03 | | | |
| Segment D (Structural Non-humanity) | 61 | 1.95d | 1.54 | | | |

Note. Segment A = low perceived non-humanity / low perceived structural similarity; Segment B = high perceived non-humanity only; Segment C = high perceived structural similarity only; Segment D = high perceived non-humanity / high perceived structural similarity. SD = standard deviation; $\eta^2$ = eta squared. Means with different superscript letters differ significantly at $p < .05$ according to Tukey's HSD post hoc test.

**Figure 3** *Mean Self-Disclosure Depth Scores across the Four Segments*

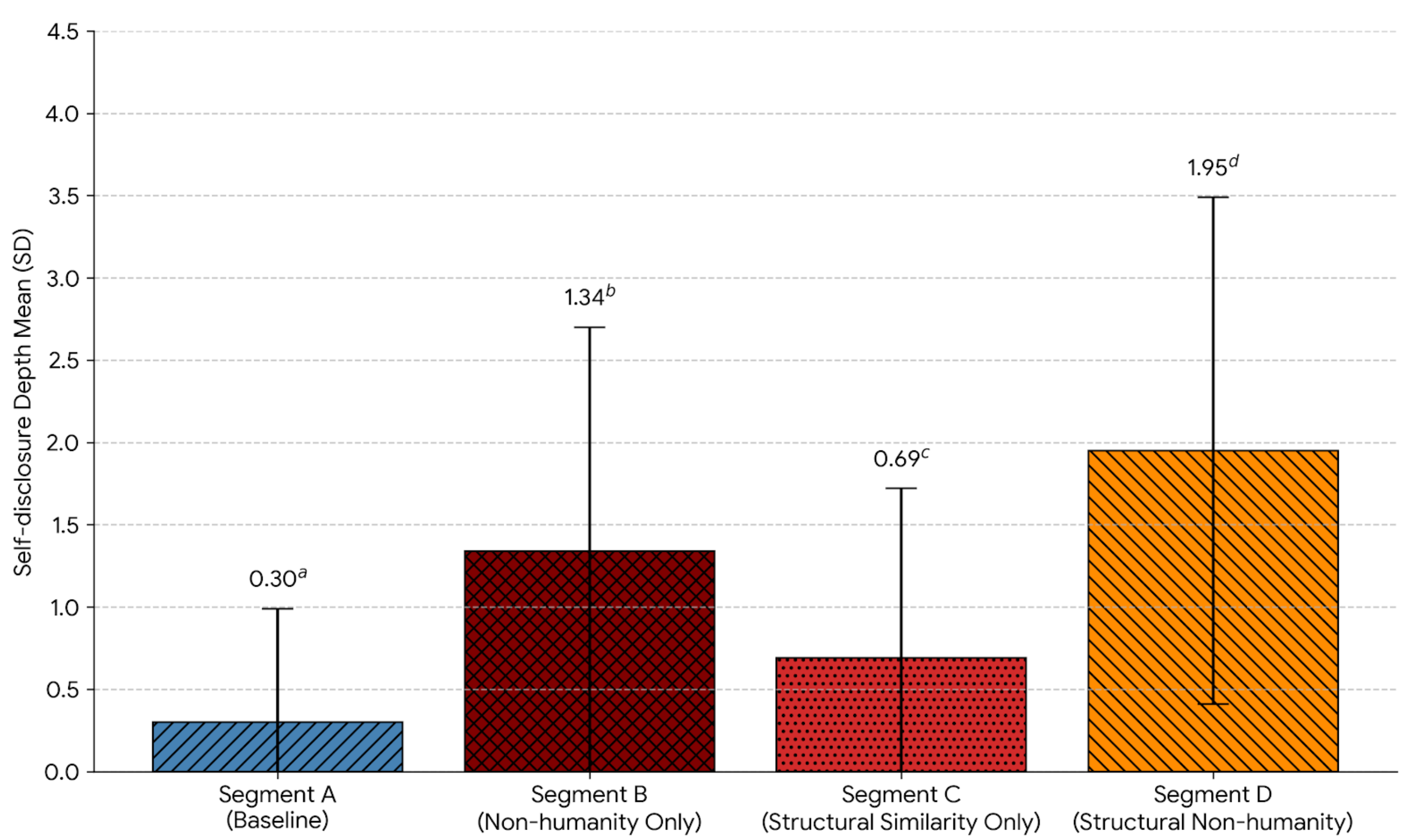

Note. Segment A = low perceived non-humanity / low perceived structural similarity; Segment B = high perceived non-humanity only; Segment C = high perceived structural similarity only; Segment D = high perceived non-humanity / high perceived structural similarity. Values indicate mean self-disclosure depth scores.

Figure 3 illustrates the mean self-disclosure depth scores (ranging from 0 to 4) across the four segments. In contrast to the Baseline (Segment A) mean of 0.30, the Structural Non-humanity (Segment D) segment exhibited a markedly higher mean of 1.95. These results suggest that the coexistence of perceived non-humanity and perceived structural similarity is strongly associated with users' engagement in deep self-disclosure.

Based on the analytical results presented above, both Hypothesis 1 (H1) and Hypothesis 2 (H2) received statistical support in this dataset.

### 4.3. Examination of Differences by Demographic Attributes (Age and Gender)

Next, to explore how the associations identified in the present model are moderated by individual attributes, a sub-segment analysis was conducted. Particular attention was given to men in their 30s and 40s, as this demographic is inferred to be more susceptible to the influence of social roles and self-inhibition.

#### 4.3.1. Tendencies among Men in their 30s and 40s

Among men in their 30s and 40s ($n$ = 438), who are assumed to exhibit higher self-inhibition due to social roles, the association between the Structural Non-humanity segment and self-disclosure was observed to be relatively more pronounced compared to other demographic groups. Within this specific demographic, the self-disclosure rate in the Baseline (Segment A) was 6.3%, whereas it reached 38.9% (OR = 9.40) in the Structural Non-humanity (Segment D) segment. Furthermore, the mean self-disclosure depth score also showed an increase from 0.21 in the Baseline segment to 1.61 in the Structural Non-humanity segment.

### 4.3.2. Tendencies among Generation Z (18–29 years old)

Among Generation Z (*n* = 389), the deep self-disclosure rate in this age group was 13.3% at the Baseline (Segment A) stage (13.3%) and reached 62.5% (OR = 10.97) in the Structural Non-humanity (Segment D). Regarding mean self-disclosure depth scores, although Generation Z exhibited a relatively higher baseline level (0.44) than other cohorts, this score reached a significantly more profound level of 2.44 in the Structural Non-humanity segment (Segment D). This pronounced difference suggests that the combined pattern proposed in the model may be especially strongly associated with AI-mediated openness among Generation Z.

#### 4.3.2.1. Liberation from the "Attention Economy" of Social Media

According to previous literature (e.g., Luo et al., 2025; Hakuhodo DY Holdings, 2025), Generation Z is susceptible to the burdens of social comparison and evaluation by others in a hyper-connected society. As a reaction to their strong tendency toward cautious self-presentation online, they may seek psychological relief in highly anonymous spaces or conversational environments with relatively low evaluation pressure. Within the present framework, the "non-humanity" factor is thought to provide a relative sense of safety—which is often difficult to achieve in interpersonal interactions—to Generation Z, whose daily lives are characterized by ubiquitous evaluation concerns.

#### 4.3.2.2. Cognitive Styles Viewing AI as an "Externalized Self"

Generation Z may be more likely to accept AI not merely as a tool, but as an auxiliary partner that extends their own cognitive processes (Chan & Lee, 2023). For these individuals, structural similarity—the degree to which the AI mirrors their own logical structure—can be naturally perceived as high consistency with their own thought processes.

### 4.3.3. Limitations of Application to the Older Population (Aged 60 and Over)

Among the demographic aged 60 and over, the number of samples corresponding to this model was small, making it difficult to provide a stable interpretation. Therefore, descriptions regarding this group should be treated as exploratory findings.

It should be noted that the results presented in this chapter describe group differences and associations based on cross-sectional survey data and do not directly indicate causal relationships.

## 4.4. Summary of Analysis Results: Dynamics of Disclosure via Structural Non-humanity

The analytical results presented in this chapter—particularly the relatively high self-disclosure rate (OR = 11.35) and depth observed in Segment D (the group perceiving both high non-humanity and high structural similarity) compared with the Baseline (Segment A)—provide important implications for understanding the factors associated with self-disclosure toward generative AI. These statistical associations can be interpreted in light of existing frameworks, namely Norman's (1988) mental models and Gentner's (1983) Structure-mapping Theory (SMT).

First, from the perspective of Norman's (1988) concept of mental models, self-disclosure may be associated with the internal predictive models that users construct through interaction with AI. The present findings suggest that the perception of AI's non-humanity is associated with lower evaluation concerns in interpersonal communication, which may in turn contribute to reducing subjective psychological barriers to self-disclosure. Ischen et al. (2023) likewise showed that the non-human nature of AI can reduce evaluation concerns and facilitate the disclosure of information with low social desirability. The current findings are broadly consistent with this line of research.

Furthermore, by applying Gentner's (1983) Structure-mapping Theory (SMT), it becomes possible to examine the consistency between the AI's responses (the system image) and the user's own thought processes (the mental model). The perception of structural similarity in AI responses can be interpreted as the establishment of a logical structure mapping between the two entities. This perception may be associated with the formation of predictability and cognitive trust toward the AI.

Taken together, the self-disclosure tendencies observed in Segment D suggest that two factors—reduced evaluation concerns through the perception of non-humanity and cognitive trust through the perception of structural similarity—may be associated with the occurrence of self-disclosure in a mutually complementary manner. These findings indicate that, in addition to the anthropomorphism-centered perspective emphasized in traditional HCI research, AI's non-humanity and the synchronization of logical structures may offer an additional perspective for understanding self-disclosure. Recent research (Hwang et al., 2025) has also qualitatively compared dialogue support provided by generative AI and humans, suggesting that AI may become a target of deep self-disclosure. Against the backdrop of these discussions, the present analysis exploratorily highlights the kinds of psychological perceptions that may be associated with self-disclosure from a cognitive perspective.

## 5. Discussion

### 5.1. Complementary Perspective to Anthropomorphism-based Models

Consistent with the hypotheses, the results suggest that deep self-disclosure to generative AI may be explained not solely by the anthropomorphism of the AI itself, but also in association with a combination of the perception of non-humanity and the perception of structural similarity. This alternative psychological pathway is consistent with the finding that the segment perceiving both high non-humanity and high structural similarity exhibited a markedly higher likelihood of disclosure (OR = 11.35) than the baseline group. This point does not entirely refute conventional anthropomorphism models; rather, it provides a complementary perspective within the context of deep self-disclosure. That is, while an AI behaving like a human may contribute to rapport or trust in some cases, in self-disclosure situations, the non-humanity of the AI may instead be linked to a reduction in evaluation concerns. The observed eleven-fold increase in the likelihood of disclosure (OR = 11.35) in Segment D, while requiring cautious interpretation, provides preliminary empirical support for this perspective.

Furthermore, a higher tendency toward deep self-disclosure was observed in the group that perceived both high non-humanity and high structural similarity. This suggests that when an AI is perceived to understand a user's flow of thought or logic to a certain degree, the perception of cognitive trust—the sense of "being understood"—may increase and may be associated with self-disclosure. Therefore, the dramatic escalation of deep self-disclosure toward generative AI observed in the present data may be related to the combination of two conditions: "not being evaluated" and "feeling understood."

However, these results describe associations based on cross-sectional survey data and do not directly indicate causal relationships. In particular, the possibility of reverse causality—where individuals with a high tendency toward self-disclosure might have retrospectively rated the structural similarity of the AI more highly—cannot be excluded. For this reason, the discussion in this section should be regarded as an exploratory interpretation of deep self-disclosure toward generative AI.

### 5.2. Variations Across Demographic Characteristics: Social Roles and Digital Nativity

The relatively strong association observed among men in their 30s and 40s provides insightful implications regarding the inhibition of self-disclosure within this demographic. For individuals in this life stage, social roles and the accompanying evaluation concerns often act as psychological barriers to openness. The findings suggest that generative AI—perceived as a non-human and non-judgmental entity—may function as a "safe target" for self-disclosure, thereby reducing the psychological costs associated with maintaining one's social persona.

Furthermore, the coexistence of perceived structural similarity and high self-disclosure in this group indicates that the AI is not merely seen as an entity to "not be cautious around," but as one that "accurately traces one’s logical flow." This combination of low evaluation risk and high cognitive understandability appears to be an important condition associated with deep self-disclosure among middle-aged men, who may prioritize logical consistency as a basis for trust.

Interestingly, similar synergistic associations were particularly pronounced among Generation Z (OR = 10.97). While this cohort inherently exhibits a higher baseline for AI-mediated openness, the notable increase observed in the Structural Non-humanity segment suggests that for digital natives, the removal of human evaluative mechanisms might further amplify their existing propensity for disclosure. As with other demographics, these findings should be interpreted as associations rather than established causal effects.

However, these interpretations remain exploratory. Possibilities remain that other factors—such as gender role norms, specific work environments, or the quality of interpersonal networks—may be involved. Therefore, these results should be understood as hypothetical findings that require further longitudinal validation to clarify the definitive psychological mechanisms across different demographics.

### 5.3. Cognitive Resolution of the Privacy Paradox

Furthermore, the present findings provide insights for understanding the "Privacy Paradox" in the field of information security. The findings suggest that the combination of perceiving structural similarity and non-humanity may help explain the cognitive dissonance between privacy concerns and the desire for openness. The observed eleven-fold increase in disclosure likelihood (OR = 11.35) suggests that this psychological framework may help explain why privacy concerns do not always prevent disclosure, even when users are aware of potential data risks.

However, this heightened tendency toward self-disclosure does not always lead to desirable outcomes. When AI facilitates the disclosure of sensitive information, ethical challenges may become more acute. Given the apparent synergy between non-humanity and structural similarity, risks such as excessive psychological dependence, unintended disclosure of sensitive information, and even potential infringements on the right to self-determination through the elicitation of disclosure warrant careful attention.

Therefore, AI design should consider not only functions that encourage disclosure but also mechanisms for the appropriate regulation of disclosure, user alerts, and transparency. For example, it may be important to implement designs that provide warnings or confirmation prompts when sensitive information is entered and clearly state that the AI is not a substitute for professional human support. Additionally, service designs that connect users to human experts in critical situations may be beneficial.

### 5.4. Social Significance and Practical Implications

These findings provide a potential basis for reconsidering interface design policies for generative AI. Rather than uniformly pursuing anthropomorphism, the results suggest that emphasizing non-evaluative attributes and logical consistency may be an effective alternative, especially for facilitating deep self-disclosure.

While the observed more than eleven-fold increase in disclosure likelihood (OR = 11.35) highlights the potential strength of this "Machine Mirror" approach, these findings should be regarded as part of an exploratory framework. Designers may consider that, for specific communicative acts, clarifying the non-human nature of the system could serve as a strategic asset in fostering a sense of psychological safety.

At the same time, given the notable escalation in disclosure rates among certain demographics, protective design principles deserve particular attention. Safeguarding users against excessive dependency should be treated as an important consideration alongside the promotion of openness.

## 6. Conclusion and Limitations

The present findings suggest, based on a secondary analysis of large-scale survey data collected in 2025, that deep self-disclosure toward generative AI may be associated with the perception of AI's non-humanity and structural similarity. In particular, the notably higher tendency observed in both the occurrence and depth of self-disclosure in the group in which both perceptions were high suggests that trust toward generative AI cannot be explained solely by the degree of anthropomorphism. Rather, it may also be understood in relation to a combination of a low perceived likelihood of being evaluated and a sense of being understood.

Several limitations should be acknowledged. First, the analysis is based on a cross-sectional survey conducted at a single point in 2025, and neither longitudinal change nor the direction of causality could be verified. Second, the primary variables were reconstructed from existing questionnaire items, and the internal consistency of the self-disclosure depth indicator was limited. Future research should develop and validate a more robust multi-item scale for disclosure depth to improve measurement reliability. Accordingly, the findings should be interpreted as exploratory. Third, because the survey sample was limited to generative AI users in Japan, the effects of cultural background and differences in usage experience could not be fully ruled out. Fourth, although the total sample size was large, the number of participants in specific segments (e.g., the Structural Non-humanity segment) and in the older population was small, which limits the generalizability of the findings to these groups. Because the focal Segment D included a relatively small number of participants, the magnitude of the observed odds ratio should be interpreted with caution and verified in future studies using more balanced group sizes. In addition, the possibility of reverse causality cannot be excluded, as individuals with a higher tendency toward self-disclosure may have retrospectively perceived greater structural similarity in the AI's responses. This possibility constitutes an important reservation in interpreting the theoretical implications of the findings.

Future research should focus on three issues. First, dedicated scales should be developed to measure the perception of non-humanity and structural similarity with greater stability and validity. Second, the causal relationship between these perceptions and self-disclosure should be tested through longitudinal and experimental studies. Third, the generalizability of the proposed framework should be examined through comparative research that takes cultural differences and variations in usage context into account.

In conclusion, the present analysis offers a theoretical perspective for reconsidering the relationship between generative AI and humans. Its validity should be further examined through cumulative future research.

**Acknowledgments**

The author would like to thank Hakuhodo DY Holdings for providing the survey data that made this research possible. During the preparation of this manuscript, the author used Gemini 1.5 Flash (Google) for English translation and structural refinement of the manuscript. The author reviewed and revised all AI-assisted output as needed, verified the accuracy of the content and references, and takes full responsibility for the content of the publication.

**Funding**

This research received no specific grant from any funding agency in the public, commercial, or not-for-profit sectors.

**Disclosure Statement**

The author reports no competing interests.

**Data Availability Statement**

The data that support the findings of this study are subject to third-party restrictions and are not publicly available. These data were provided by Hakuhodo DY Holdings and may be available from the author upon reasonable request and with permission from Hakuhodo DY Holdings.